\documentclass[aps,showpacs,prl,twocolumn]{revtex4-1}
\usepackage{dcolumn}
\usepackage{bm}
\usepackage{amssymb}
\usepackage{amsmath}
\usepackage{amsfonts}
\usepackage[]{graphicx}
\usepackage{comment}
\usepackage{color}

\begin{document}

\title{The acoustic radiation force: a gravitation-like field.}
\author{ Pierre-Yves Gires$^{1}$, Jerome Duplat$^{2}$, Aur\'{e}lien Drezet$^{3}$ \& Cedric Poulain$^{1,3}$}
\affiliation{(1) Univ. Grenoble Alpes, CEA, LETI,  F-38000 Grenoble, France}
\affiliation{(2) Univ. Grenoble Alpes, INAC-SBT, F-38000 Grenoble, France}
\affiliation{(3) Univ. Grenoble Alpes, CNRS, Grenoble INP, Institut Neel, F-38000 Grenoble, France}

\begin{abstract}

In this letter, we propose an expression for  the \emph{instantaneous}  acoustic radiation force acting on a compressible sphere
when it is  immersed in a sound field  with a wavelength much larger than the particle size  (Rayleigh scattering  regime).
We show that the leading term of the
radiation force can alternatively be expressed as the time average of a fluctuating gravitation-like
force. In other words, the  effect of the acoustic pressure gradient 
is to generate a local acceleration field
encompassing the sphere,
which gives  rise to an apparent 
buoyancy force, making the object move in the incoming field.
When averaging over time, we recover the celebrated  Gor'kov expression and emphasize that
two terms appear, one local and one convective,
which identify  to the well-known monopolar and dipolar contributions.
\end{abstract}

 \maketitle

Since Rayleigh's pioneering work on sound waves \cite{Rayleigh1902}, later followed by Langevin and Brillouin,
among others \cite{Chu1982, Mcintyre1981, Brillouin1925}, it is known that like its electromagnetic cousin, acoustic waves can transfer linear momentum to a particle, even in a perfect fluid.
The mean effect is referred to as the radiation or pressure force, with its associated tension or radiation stress tensor \cite{Brillouin1925}.
Upon reflection, one might wonder how an elemental sound wave, with its harmonic pressure-velocity oscillation (in time and space), could exert a non-zero average force.
Like with electromagnetic waves, the answer lies in the second order effect arising for sound waves: the particle pulsates in volume while moving back and forth in the acoustic oscillating flow, yielding a small hysteretic displacement. 
These incremental displacements accumulate over millions of cycles per second (at the sound frequency), and may be interpreted as the result of an average or macroscopic force on the particle.

Such an average force was first
calculated by King for a hard  sphere in a perfect fluid~\cite{King1934}, and generalized by Yosioka~\cite{Yosioka1955} for compressible objects.
 It still remains an active  subject of research
 with increasingly heavy mathematics
addressing complex objects, wave fields, and more realistic effects \cite{Johnson2016, Baresch2013}.
Likewise, impressive applications  of the radiation  force 
to microfluidics have been published  in the last ten years.  This has given rise to the new discipline of acoustofluidics \cite{bruus2011LOCtutorials}, in which the ability of standing waves to arrange, trap and sort living cells has been demonstrated both in propagative \cite{Augustsson2016} and evanescent fields \cite{Aubert2016}.

Despite this  great success
in applications, very few papers address
fundamental questions such as the effect of 
added mass or compressibility at short time when a sphere
is accelerated by the sound field and becomes non-buoyant in the fluid.

In order to gain insight into the physics at play, here we will focus on the instantaneous (i.e., not averaged) dynamics
of the particle in the incoming field.
By following a Lagrangian approach
and comparing with the celebrated 
Gor'kov expression of the radiation force \cite{Gorkov1962},
 we will show that the  radiation force  can  also 
be interpreted as the average of a fluctuating buoyant force,
shedding new light on its physical origin. 

Let us consider a non-moving, infinite and  non viscous compressible fluid
in which a compressible
sphere of radius $a$ is immersed, in the absence of gravity.
The whole system is excited by a time-harmonic acoustic wave 
characterized by its incident Eulerian velocity, pressure and density fields, respectively 
$(\boldsymbol{v}_{\textrm{in}},~p_{\textrm{in}},~\varrho_{\textrm{in}})$.
The acoustic wavelength $\lambda_{f}$ in the fluid
is supposed much larger than the sphere radius (Rayleigh regime): $a \ll \lambda_{f}$.
Assuming that after a certain time a steady state  can be reached \cite{Beissner1998}, 
we define the radiation force $\bar{\mathbf{F}}_\text{rad}(\mathbf{r})$ as the time-averaged force exerted on any particle positioned at position $\mathbf{r}$.
\begin{eqnarray}
\bar{\mathbf{F}}_\text{rad}(\mathbf{r})=\langle\mathbf{F}(\mathbf{r}_{p}(t))\rangle,
\label{eq:rad_force_def}
\end{eqnarray}
where $\mathbf{F}(\mathbf{r}_{p}(t))$ is 
the instantaneous force exerted upon the small particle when the medium is insonified 
and  $\mathbf{r}_{p}(t)$ the positionof the particle at time $t$.

In an inviscid fluid, only pressure forces can apply on the 
particle, so that
\begin{equation}
\mathbf{F}(\mathbf{r}_{p}(t))=\int_{S_p(t)}p\boldsymbol{dS},\label{eq:f_fl_part_inst}
\end{equation}
$p$ being the total pressure field (incident and scattered), the normal being oriented 
towards the particle surface $S_p(t)$.
For concision, $\mathbf{F}(\mathbf{r}_{p}(t))$ will be hereafter denoted $\mathbf{F}(t)$, keeping in mind that it refers to the moving particle located at position  $\mathbf{r}_{p}(t)$.
In general, the instantaneous force
$\mathbf{F}(t)$ is not known,  only its averaged value.

By using an asymptotic approach detailed below,  
we are going to derive the expression for an equivalent instantaneous radiation force $\mathbf{F}_{\text{rad}}(t)$, so that
at the leading order in the Mach number, $\mathbf{F}(t)=\mathbf{F}_{\text{rad}}(t)+\mathbf{f}(t)$,
with $\mathbf{f}$ a zero-mean function.

The (small) acoustic Mach number $\varepsilon$ is: 
$\varepsilon=\frac{v_{in}}{c_{f_{0}}}$,
 $\small{c_{f_{0}}=1/\sqrt{\varrho_{f0}\kappa_{f0}}}$ being the fluid sound velocity, with respectively $\varrho_{f0} \mbox{ and } \kappa_{f0}$ its equilibrium density and compressibility, while $v_{in,c}$ is the characteristic fluid velocity.
The subscripts $p$ and $f$ refer respectively to the particle and the fluid, while the $0$ subscript denotes quantities at rest, \textit{i.e.} in the absence of acoustic perturbation.
Following  Xie \textit{et al.} \cite{Xie2014}, we also describe the scaling of the particle size with an exponent $\alpha$, so that the particle radius $a$ over the wavelength $\lambda$ is such that
\begin{equation}
\frac{a}{\lambda}=O(\varepsilon^\alpha).
\label{eq:alpha_def}
\end{equation}

In this framework, a well known approach for estimating  the 
mean radiation force 
$\bar{\mathbf{F}}_{\text{rad}}$
was undertaken by Gor'kov \cite{Gorkov1962} and was subsequently detailed by Bruus \cite{Bruus2012b, Settnes2012}.
It consists in transforming the integral over the time varying surface particle $S_p(t)$ into  an integral over a  fixed remote surface where the far-field approximation allows to find  the leading term.

We begin by recalling the Gor'kov approach, on which our analysis is partly based. Some 
implicit assumptions of the theory are discussed in the Supplemental Material (SM) 1~\cite{Supplemental}.
Introducing  $\boldsymbol{\Pi}$ the total momentum density flux tensor defined as $\Pi_{ij}=p\delta_{ij}+\varrho v_i v_j$, 
a momentum balance shows that the radiation force can be approximated by the mean momentum flux through any fixed surrounding surface $S$
\begin{equation}
\bar{\mathbf{F}}_{\text{rad}}\simeq  -\int_{S}\langle\boldsymbol{\Pi}\rangle\cdot\boldsymbol{dS},
\end{equation}
with $\boldsymbol{dS}$ oriented outward the surface $S$.  
Here, as done by Gor'kov and \cite{Bruus2012b}, the  additional contribution $\langle\frac{d}{dt}\int_{V(t)}\varrho\boldsymbol{v}dV\rangle$ associated  with the  rate of linear momentum of the fluid located between the surfaces $S$ and $S_p(t)$  has been omitted (see SM 1~\cite{Supplemental} for a discussion).
Then, choosing $S$ in the far field region, recognizing that the leading term in $\langle\boldsymbol{\Pi}\rangle$ only depends on the particle monopolar and dipolar contributions, Gor'kov shows that for a standing incident field the leading term of $\bar{\mathbf{F}}_{\text{rad}}$ is $\rho c^2 a^2 O( \varepsilon^{2+\alpha})$ and derives from the acoustic potential $U(\mathbf{r})$ so that

\begin{equation}
\bar{\mathbf{F}}_{\text{rad}}\simeq-\boldsymbol{\nabla}U(\mathbf{r})
 \label{eq:rad_force},
\end{equation} 
with 
\begin{eqnarray}
U(\mathbf{r})&=&V_{p0}\left(\frac{f_1}{2}\kappa_{f0}\langle p^2_{\textrm{in}}\rangle-\frac{3f_2}{4}\varrho_{f0}\langle v^2_{\textrm{in}}\rangle\right)\label{eq:Gor'kov_pot}\\
f_1&=&1-\tilde{\kappa}\label{eq:f1_def},\\
f_2&=&\frac{2(\tilde{\varrho}-1)}{2\tilde{\varrho}+1}\label{eq:f2_def},
\end{eqnarray}
$\left(\tilde{\kappa}=\frac{\kappa_{p0}}{\kappa_{f0}},\tilde{\varrho}=\frac{\varrho_{p0}}{\varrho_{f0}}\right)$ 
being respectively the equilibrium compressibility and density ratios of the particle over the fluid, while $V_{p0}$ is the particle rest volume.

With this in mind let us now go back to  the definition of the scattered pressure field $p_{s}$
as the correction  of the incident field 
required to account for the presence of the particle:
\begin{equation}
p= p_{\textrm{in}}+ p_{s}.
\end{equation}
From Eqs.~\ref{eq:rad_force_def} and \ref{eq:f_fl_part_inst} it is thus always possible to calculate  $\bar{\mathbf{F}}_{\text{rad}}$ as $\bar{\mathbf{F}}_\text{rad}(\mathbf{r})=\langle
\int_{S_p(t)}(p_{\textrm{in}}+ p_{s})\boldsymbol{dS}\rangle$. This was the approach used in the seminal work of Yosioka~\cite{Yosioka1955} in which an expansion equivalent to the one used in optics by Mie for scattering of light  by spherical particle was done to deduce Eq.~\ref{eq:rad_force}. 

Here however, we  instead consider a Lagrangian description of the fluid motion surrounding the moving particle. For this purpose, we first make a guess that the motion of the particle is mainly driven by the effect of $p_{\textrm{in}}$. As a first approximation we could be tempted to write $\bar{\mathbf{F}}_\text{rad}(t)=
\simeq \langle
\int_{S_{p,in}(t)}p_{\textrm{in}}\boldsymbol{dS}\rangle$ where $S_{p,in}(t)$  is the surface of the particle  altered  only by the effect of $p_{\textrm{in}}$ (see SM 1 \cite{Supplemental} for a rigorous definition). However, this approximation is in general too crude. Indeed, in the limit case of a fluid particle in fluid, i.e. with $f_1=f_2=0$, we get from Eqs.~\ref{eq:rad_force} and \ref{eq:f2_def} that $\bar{\mathbf{F}}_{\text{rad}}=\mathbf{0}$. This contradicts the elemental fact that, in such a case, for a standing wave, $\langle
\int_{S_{p,in}(t)}p_{\textrm{in}}\boldsymbol{dS}\rangle$ doesn't actually vanish (See SM 3 \cite{Supplemental}). Therefore, the expression should be corrected, the simplest one we have found being of the form     

\begin{eqnarray}
\bar{\mathbf{F}}_\text{rad}\simeq-\langle \beta(t)
\int_{S_{p,in}(t)}p_{\textrm{in}}\boldsymbol{dS}\rangle,\label{eq:def_fa_fb}
\end{eqnarray}
where the specific correction
$\beta(t)= \frac{\varrho_{p}-\varrho_{f}}{\varrho_{f0}}$, $\varrho_p(t)$ and $\varrho_f(t)$ being respectively the instantaneous particle and fluid densities in the mere incident field, has the peculiar property to cancel at all times for a neutral particle (\textit{i.e.} when no radiation force is present).
Remarkably, we find that this choice for $\beta(t)$ allows us to recover Gor'kov results; i.e.,  Eq.~\ref{eq:rad_force}.

To see that, first, let us rewrite Eq. \ref{eq:def_fa_fb}  as $\bar{\mathbf{F}}_\text{rad}(t)\simeq\langle\mathbf{F}_a(t)\rangle $ where 
\begin{eqnarray}
\mathbf{F}_{a}(t)&=&-\frac{(\varrho_{p}-\varrho_{f})}{\varrho_{f0}} \int_{S_{p,in}(t)}p_{\textrm{in}}\boldsymbol{dS}\label{eq:inc_press_term_0}\\
&=&\frac{(\varrho_{p}-\varrho_{f})}{\varrho_{f0}}\boldsymbol{\nabla}p_{\textrm{\textrm{in}}}(\mathbf{r}_p(t))V_{p,in}(t).\label{eq:inc_press_term_2}
\end{eqnarray}
Then, in order to calculate  $\langle\mathbf{F}_{a}(t)\rangle$,
one must remember that the particle  constantly moves in the field.
To simplify the present calculations, we can define $\tilde{\mathbf{F}}_a(\mathbf{r},t)$, obtained from the expression of $\mathbf{F}_a(t)$ assuming the particle stands at $\mathbf{r}$ instead of $\mathbf{r}_p(t)$ at time $t$, thus satisfying 

\begin{equation}
\mathbf{F}_a(t)=\tilde{\mathbf{F}}_a(\mathbf{r}_p(t),t).
\end{equation}

As $\mathbf{r}_p(\boldsymbol{0})=\mathbf{r}$, we then have
\begin{equation}
\small{\mathbf{F}_{a}(t)\simeq\mathbf{\tilde{F}}_a(\mathbf{r},t)+\Big(\mathbf{r}_{p}(t)-\mathbf{r}_{p}(0)\Big)\cdot\boldsymbol{\nabla}\mathbf{\tilde{F}}_a(\mathbf{r},t)}\label{eq:taylor_expansion}.
\end{equation}

We identify both a \emph{local} and \emph{convective} contributions, respectively denoted
 $\mathbf{F}_a^{\text{loc}}(t)$ and $\mathbf{F}_a^{\text{conv}}(t)$, so that
\begin{equation}
\mathbf{F}_a(t)\simeq\mathbf{F}_a^{\text{loc}}(t)+\mathbf{F}_a^{\text{conv}}(t),
\end{equation}  
with
\begin{eqnarray}
\mathbf{F}_a^{\text{loc}}(t)=\mathbf{\tilde{F}}_a\left(\mathbf{r},t\right),\label{eq:loc_term}
\end{eqnarray}
and
\begin{eqnarray}
\mathbf{F}_a^{\text{conv}}(t)=\int_{0}^t\boldsymbol{v}_{p}(t')dt'\cdot\boldsymbol{\nabla}\mathbf{\tilde{F}}_a\left(\mathbf{r},t\right).\label{eq:conv_term}
\end{eqnarray}

The local force can be expressed, keeping terms up to order $O(\varepsilon^{2+\alpha})$, 
as
\begin{equation}
\small{\mathbf{F}_a^{\text{loc}}(t)=-V_{p0}\left[-\frac{3}{2}\frac{f_2}{1-f_2}\boldsymbol{\nabla}p_{\textrm{in}}+\kappa_{f0}\frac{f_1}{2} \boldsymbol{\nabla}p^2_{\textrm{in}}\right](\mathbf{r},t)}
\label{eq:loc_term_exp},
\end{equation}
where use of  Euler's equation has been made as detailed in SM 2.

Averaging  the above expression over time then yields
\begin{equation}
\langle\mathbf{F}_a^{\text{loc}}(\boldsymbol{r_{p}}(t))\rangle=-\boldsymbol{\nabla} \left[V_{p0}\kappa_{f0}\frac{f_1}{2} \langle p^2_{\textrm{in}}\rangle \right].
\label{eq:loc_term_mean}
\end{equation}
It is noteworthy that the term in brackets
 corresponds to the first term (monopolar contribution)  in the Gor'kov  force potential  expressed in Eq. \ref{eq:Gor'kov_pot}.

In order to obtain the mean convective term contribution, we start expressing the first order particle velocity $\boldsymbol{v}_{p}(t)$ as a function of the incident acoustic velocity field 
$\boldsymbol{v}_{\textrm{in}}$.
The particle being accelerated in the incident sound field of velocity $\boldsymbol{v}_{\textrm{in}}(\mathbf{r},t)$,
the surrounding fluid inertia leads to an added mass effect (see \cite{Landau1987}),
which first appears at the order
$O(\varepsilon)$:
At this order,  Batchelor \cite{Batchelor1970}
 shows that  $\boldsymbol{v}_{p}$ can be expressed as
 
\begin{eqnarray}
\boldsymbol{\dot{r}}_{p}(t) = \boldsymbol{v}_{p}(t)&=&(1-f_2)\boldsymbol{v}_{\textrm{\textrm{in}}}(\bold{r}_{p}(t))\label{eq:vel_first_order}.
\end{eqnarray}

Now,  inserting expressions \ref{eq:loc_term_exp} and \ref{eq:vel_first_order} in Eq. \ref{eq:conv_term}, we get
\begin{equation}
\mathbf{F}_a^{\text{conv}}(t)=V_{p_0}\frac{3}{2}f_{2}\int_{0}^{t}\boldsymbol{v}_{\textrm{in}}(\bold{r},t')dt'\cdot\boldsymbol{\nabla}\left(\boldsymbol{\nabla}p_{\textrm{in}}(\bold{r},t)\right)\label{eq:conv_term_dev1}.
\end{equation}

After some calculations  detailed in SM 3, involving the linearized Euler equation and an integration by parts,
it comes
\begin{equation}
\langle\mathbf{F}_a^{\text{conv}}(\mathbf{r}_{p}(t))\rangle=-\boldsymbol{\nabla} \left [ -V_{p0}\frac{3f_2}{4}\varrho_{f0}\langle v_{\textrm{in}}^2\rangle \right ].
\label{eq:conv_term_mean}
\end{equation}

\begin{figure*}
\begin{minipage}{0.49\textwidth}
\flushleft{a)}
\includegraphics[width=\linewidth]{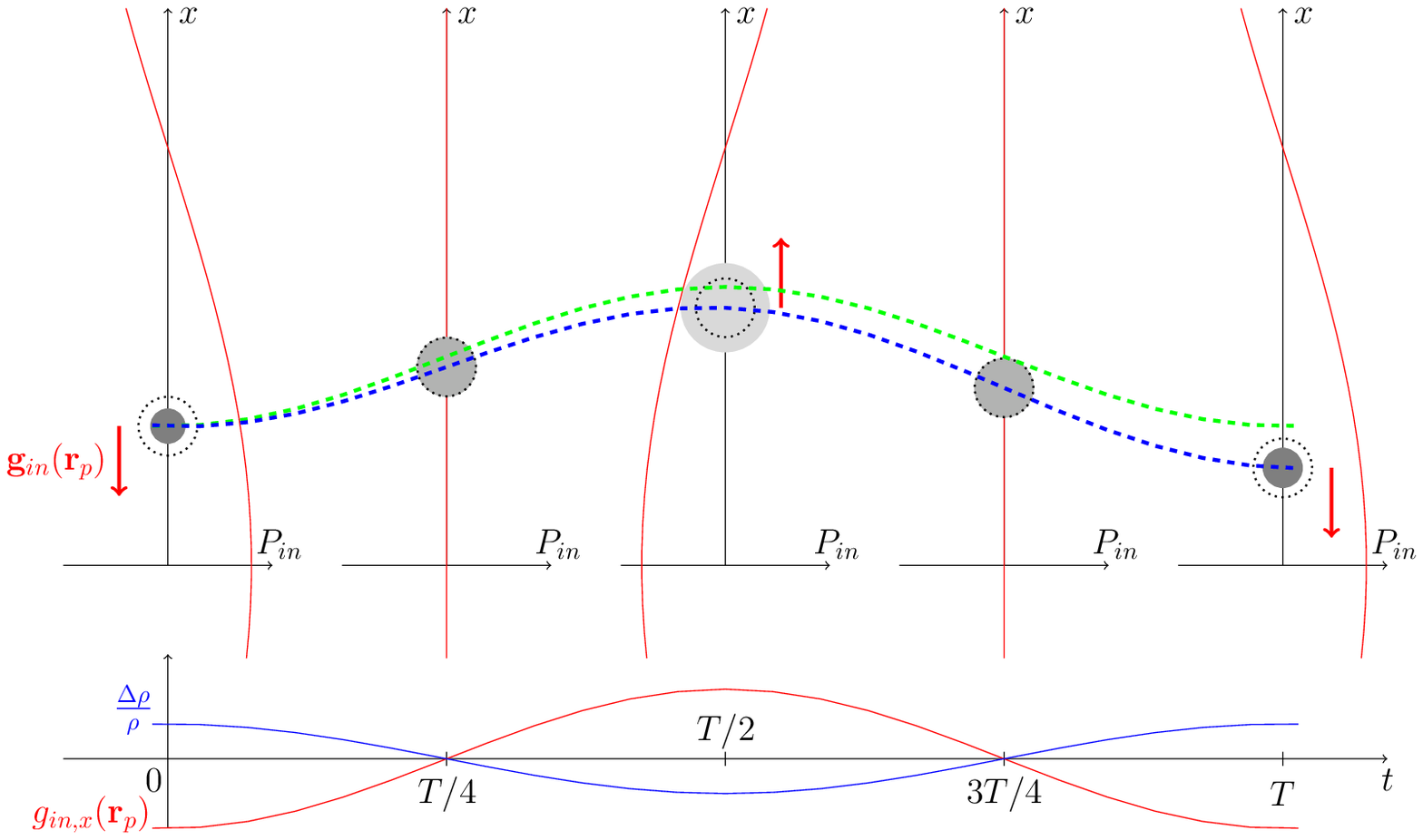} 
\end{minipage}
\begin{minipage}{0.49\linewidth}
\flushleft{b)}
\includegraphics[width=\linewidth]{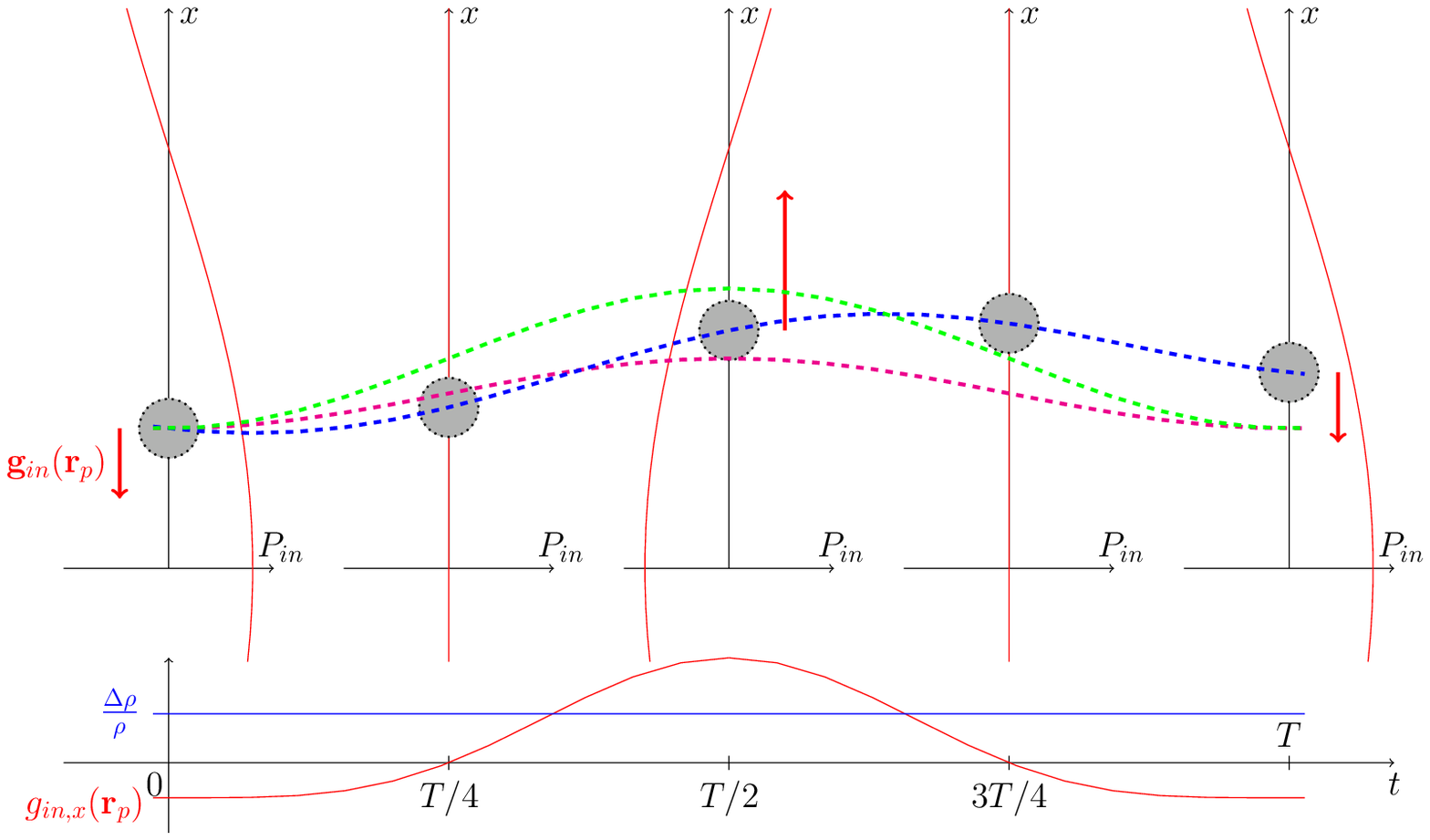} 
\end{minipage}
\caption{
Time sequence  of a particle 
 over one period  $T$ of  a standing  wave
 propagating along the (vertical) $x$ axis.
It illustrates the radiation force as a gravity like effect 
for two representative cases
plotted according to Eq.  \ref{eq:flottability}:
(a) a compressible and neutrally buoyant particle 
and (b)  a dense and iso-compressible particle
The green dashed line is the trajectory the particle would follow as a fluid particle, and the blue dashed  line is the actual one. 
On case (b) is also added in magenta the trajectory the particle would have without the added-mass effect.
The sphere volume is filled with grey color, its equilibrium shape being delimited with the black dotted line. As it plays no role in the second case, we chose to represent an incompressible particle.
Below each sequence, both the local relative density shift $\Delta \varrho/\varrho$ and  the acoustic gravity component $g_{in,x}$ are plotted to ease interpretation.
See the text for a more detailed step by step explanation.
}
\label{fig:acoustic_gravity}
\end{figure*}
Likewise, the term in brackets is the dipolar contribution
of the Gor'kov potential expressed in Eq. \ref{eq:Gor'kov_pot}.

Combining the mean local and convective contributions given in Eqs. \ref{eq:loc_term_mean} and \ref{eq:conv_term_mean}, 
we obtain, at the leading order,
\begin{equation}
\label{eq:tot_cont}
\bar{\mathbf{F}}_{\text{rad}}\simeq\Big\langle\mathbf{F}_{a}\left(\mathbf{r}_p(t)\right)\Big\rangle.
\end{equation}

$\mathbf{F}_{a}(t)$ can thus be identified to the instantaneous radiation force $\mathbf{F}_{\text{rad}}(t)$ defined in Eq. \ref{eq:rad_force} in complete agreement with Gor'kov's results. This Lagrangian derivation of the radiation force constitutes the first main finding of this letter. The second important result, that we will now discuss, concerns the physical interpretation of this radiation force.   

For this purpose we introduce an effective gravitation field $\boldsymbol{g}_{\textrm{in}}=\frac{\boldsymbol{\nabla}p_{\textrm{in}}}{\varrho_{f0}}$ and a relative density $\Delta \varrho=\varrho_p-\varrho_f$ in Eq. \ref{eq:inc_press_term_2} and Eq. \ref{eq:tot_cont}, which leads to
\begin{equation} 
\label{eq:flottability}
\mathbf{\bar{F}}_{\text{rad}}= \langle \Delta \varrho V_{p} \boldsymbol{g}_{\textrm{in}} \rangle.
\end{equation}

The radiation force can thus be understood  as the leading term in the mean force which would result from a gravitational field modulation equal to 
$\boldsymbol{g}_{\textrm{in}}(t)$.
In other words,  $\mathbf{F}_{a}(t)$ can be seen as a \emph{fluctuating} apparent weight, resulting 
from the combination of two 
oscillating quantities:

\begin{itemize}
\item a forcing effect: an incident acoustic gravity-like acceleration field $\boldsymbol{g}_{\textrm{in}}(\mathbf{r},t)$
\item a response effect: owing to their compressibility, both the particle volume and the fluid densities 
oscillate
at the forcing frequency, rendered by the relative density term 
$ \Delta \varrho  V_{p}=\frac{\Delta \varrho}{\varrho_p}m_p$ ($m_p$: constant particle mass) 
\end{itemize}

By analogy with the buoyancy force (i.e. Archimede's law)
arising when a particle has a density or compressibility different from the fluid in which it is immersed,
the oscillating force $\mathbf{F}_{a}(t)$  is equivalent to a rapidly fluctuating  `acoustic gravitational force'.
As we have shown, the time-average of this force, taking both the temporal and spatial structure of the field into account,
leads to the classic radiation force expressed  by Gor'kov
for standing waves.
Our conclusion is also in agreement with Gor'kov's work
for progressive wave for which
the radiation force is expected to  be  zero at this order  as well.

We will now try to give a comprehensive  picture of the physics
at play in the gravitation-like force, considering  a particle in a plane standing wave $p_{\textrm{in}}=-p_0\cos{(\omega t)}\cos{(kx)}$, with
$k=2\pi/\lambda$ and $\omega=k c_{f0}$ respectively the wavenumber and angular frequency of the wave.
As explained above,
the acoustic gravitational effect results from two contributions: a local one, associated with the oscillation of the particle apparent mass $m_p(1-\frac{\varrho_{f}}{\varrho_p})$ in the acoustic gravitation field, and a convective one, linked to
the local exploration of the field by the oscillating particle. 
Let us now separate both contributions by considering
two limit cases for a particle initially located between a pressure antinode (at $x=0$) and the nearest node in the $x>0$ region
(see Fig. \ref{fig:acoustic_gravity}).

\paragraph*{\textbf{Case a:} a neutrally buoyant but compressible particle}

We first consider a particle both neutrally buoyant 
($\varrho_{p0}=\varrho_{f0}$)  and more compressible than the fluid  ($\kappa_{p0}>\kappa_{f0}$, \textit{i.e.} $f_1< 0$). 
In this case, only the local contribution remains as the convective term vanishes ($f_{2}=0$).
Let us figure out  the particle movement over a time period $T$.
The sinusoidal green dashed line on  Fig. \ref{fig:acoustic_gravity}a) represents the movement of a \emph{fluid} particle in the sound wave, which is also, 
at leading order,
the particle movement
since $f_{2}=0$  so that $\boldsymbol{v}_{p}=(1-f_{2})\boldsymbol{v}_{\textrm{in}}=\boldsymbol{v}_{\textrm{in}}$ (no added mass effect comes into play in this case).

At time $t=0$, the pressure gradient is such that $\boldsymbol{g}_{\textrm{in}}$ is oriented downward
\footnote{For the sake of conciseness, we will abusively use the terms up (towards $x  > 0$)
and down (towards $ x  < 0$), assuming the classical paper reading orientation on Earth.}.
The particle is compressed and hence denser than the hosting fluid at its location, so that it 
plunges downwards.
At $t=T/4$, the pressure is zero, the particle thus following the non perturbed trajectory (no radiation force).
At $t= T/2$, the pressure gravity reverses but the particle also expands, so that it is `lighter' and thus keeps sinking.
At $t=3T/4$ it follows the same trajectory (no force).
Overall, the acoustic gravity is always out of phase with the density shift:  for the whole period,
the particle keeps  `falling' towards the pressure antinode at $x=0$
and the instantaneous radiation force maintains the same orientation.

\paragraph*{\textbf{Case b:} an iso-compressible particle, denser than the fluid}

We now consider a particle iso-compressible ($f_1=0$) but denser than the hosting fluid  ($f_2>0$), also  located between $x=0$ and $x=\pi/k$, as shown on Fig. \ref{fig:acoustic_gravity}b). 
To the green fluid particle trajectory is now added the flattened magenta dashed line, which represents the trajectory the particle would follow if only added-mass effect would apply.
First, the particle dynamic is such that it
plunges between $t=0$ and $t=T/4$ since it is denser than the fluid in a downward gravity field.
However, at $t=T/2$, the gravity field reverses so that the particle rises.
Here, the point is that the gravity field (or acoustic pressure gradient) being greater at the location where the particle
is at $T/2$ than at $t=0$, the radiation force over the cycle does not balance symmetrically.
On average, a net upward radiation force (toward the nearest node) is exerted upon the particle.

It is noteworthy that in this case, the effect of the added mass 
is to alter  (with a factor $(1-f_{2})$) the particle trajectory obtained by integrating $\boldsymbol{v}_{p}=(1-f_{2})\boldsymbol{v}_{\textrm{in}}$.
In other words, the `landscape' 
explored by the particle determines the 
hysteretic  values
of the gravity field that the particle will encounter at extremal positions
and henceforth contributes to its amplitude.
In the convective (or `landscape')  effect, we note that the radiation force reverts within a cycle 
while it is not the case in the local one, where it maintains the same orientation.

In conclusion, the time-resolved Lagrangian approach is another way
to illustrate  both the  time-averaged scattered contributions (monopolar and dipolar scattering terms)
appearing in the Gor'kov potential of the averaged radiation force.
Eq.~\ref{eq:flottability} gives a clear interpretation  (see Fig. \ref{fig:acoustic_gravity})  of the acoustically induced buoyancy effect, from which both terms are derived:
(i) a local term, originating from  the compressibility ratio, a neutrally buoyant particle alternatively sinking and floating within the time varying acoustic field, 
and (ii) a convective one, a denser but iso-compressible particle encountering a stronger instantaneous force as it approaches pressure nodes where the pressure gradient is the largest.
Overall, it shows that the radiation force has the structure of an inertial force in an
\emph{equivalent}
gravity acceleration field created by the acoustic pressure gradients in the fluid.
Therefore, it is possible to consider the acoustic radiation force as a linear centrifugal force.
This idea is indeed reinforced by 
the ability of the acoustic  radiation  force to sort  particles according to their size, density and compressibility ratios (acoustophoresis \cite{bruus2011LOCtutorials}), which is reminiscent  of  centrifugal forces.
Correlatively, our findings  may also 
shed new light on the ability of the radiation force to universally deform interfaces \cite{Wunenburger2012}, or  to separate  miscible fluids of different densities, as  recently evidenced \cite{Karlsen2016, Karlsen2017}.
Beyond acoustics, we wonder  
how this approach could be transposed to other wave natures 
for which a radiation force also arises.
We hope this paper will stimulate future work in this direction.

\bibliography{biblio_radiation_force_2018}
\end{document}


\begin{center}
\textbf{\large Supplemental Material}
\end{center}
\setcounter{equation}{0}
\setcounter{figure}{0}
\setcounter{table}{0}
\setcounter{page}{1}
\makeatletter
\renewcommand{\theequation}{S\arabic{equation}}
\renewcommand{\thefigure}{S\arabic{figure}}

\section{SM1. Mean momentum balance and the Gor'kov approximation}

\label{app:mean_mom_balance}

\begin{figure}[h!]
\centering
\includegraphics[width=\linewidth]{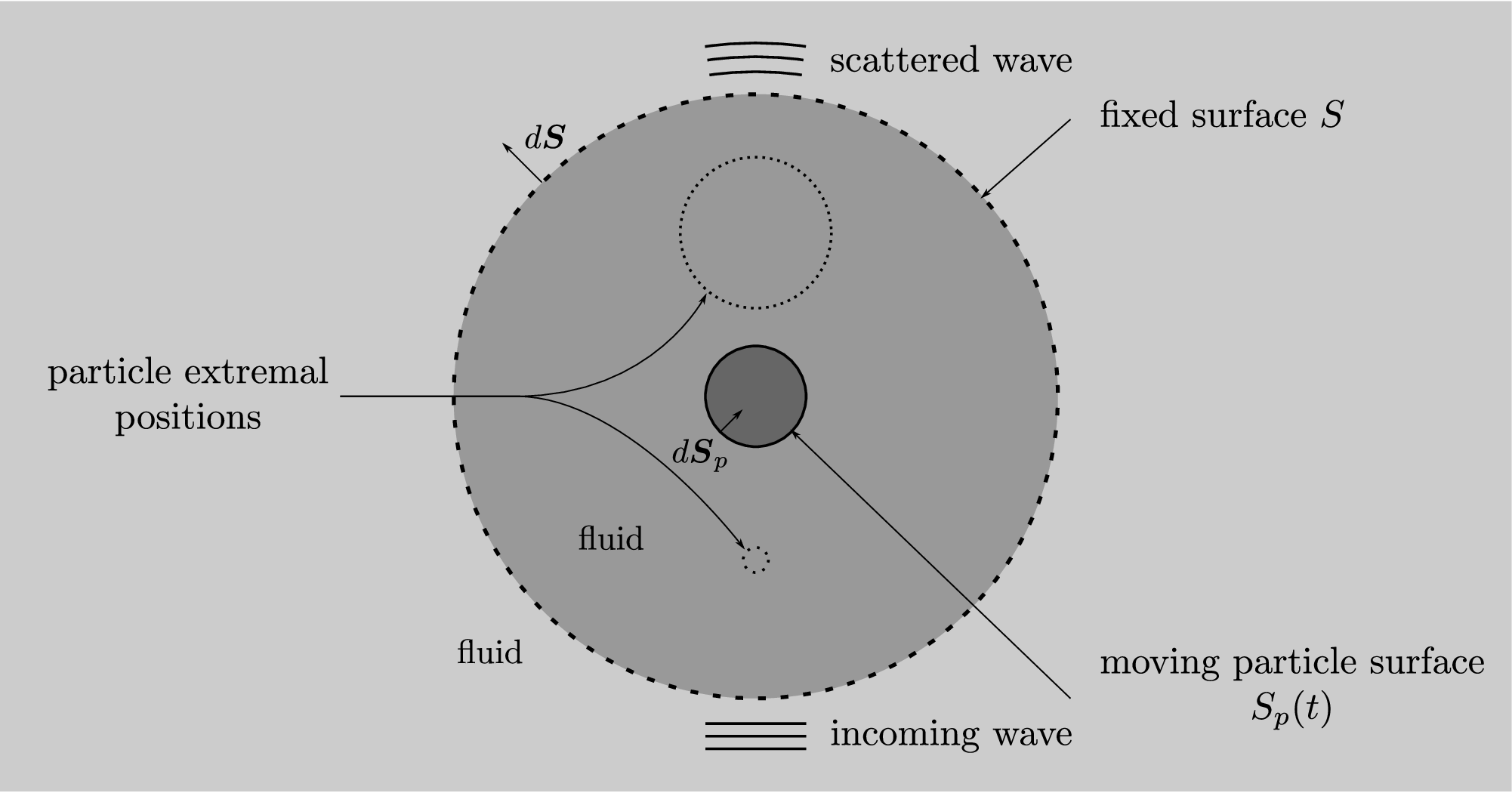} 
\caption{Notations used for the momentum balance. \label{fig:geom_def}
}
\end{figure}

In this section, we are going to discuss the applicability  of the Gor'kov
formulation (see \cite{Gorkov1962}) in the context of a free-to-move (not suspended) particle.
It is to our knowledge an issue which has never been discussed in details in the literature so far although
it is mostly employed in this context.

Let us consider a compressible particle
of varying outer surface $S_p(t)$
immersed in a sound field, and define a much larger surface $S$
in the far-field region.
$V(t)$ is  the  control volume $V$ delimited externally by $S$ and internally by $S_p(t)$, as sketched on Fig. \ref{fig:geom_def}. 

In the general case, the momentum change  rate of the fluid volume $V$ writes
\begin{equation}
\label{eq:mom_bal_4}
\frac{d}{dt}\int_{V(t)}\varrho\mathbf{v}dV=-\int_{S}\boldsymbol{\Pi}\cdot\boldsymbol{dS}-\int_{S_p}p\boldsymbol{dS}.
\end{equation}

Averaging Eq. \ref{eq:mom_bal_4} over time, we then get
\begin{eqnarray}
\bar{\mathbf{F}}_{\text{rad}}&=&-\langle\frac{d}{dt}\int_{V(t)}\varrho\mathbf{v}dV\rangle-\int_{S}\langle\boldsymbol{\Pi}\rangle\cdot\boldsymbol{dS}\\
&=&-\int_{S}\langle\boldsymbol{\Pi}\rangle\cdot\boldsymbol{dS}-\frac{\boldsymbol{P}(T)-\boldsymbol{P}(0)}{T}\label{eq:momentum_balance_gen},
\end{eqnarray}
with
\begin{equation}
\boldsymbol{P}(t)=\int_{V(t)}(\varrho\mathbf{v})(t)dV.
\end{equation}

In order to  simplify the problem,
let us first assume that the  movement of the particle mass center    in the sound flow is perfectly periodic in space and time so that the particle mean displacement is exactly zero.
This can be achieved by means of an additional external and time independent force  hereafter denoted $
\mathbf{F^*}$.
Hence, we have
\begin{equation}
\mathbf{F^*}+\int_{S_p(t)} p \mathbf{n}  dS = m_p \mathbf{\Gamma}(t),
\end{equation}
 with $\mathbf{\Gamma}$ and $m_p$ respectively the particle acceleration and mass.

By averaging over time,  it is clear that  the force $\mathbf{F^*}$ must exactly compensate the mean radiation force $\bar{\mathbf{F}}_{\text{rad}}=\langle \int_{S_p} p \mathbf{n} dS \rangle$ to ensure a zero-mean displacement (or acceleration).

Thanks to the additional force, the movement is perfectly periodic in time, so that the term $\langle\frac{d}{dt}\int_{V(t)}\varrho\mathbf{v}dV\rangle $ cancels.
With this assumption of a \textit{suspended} particle,
we recover the expression for the radiation force as used by Gor'kov (first equation of \cite{Gorkov1962}), also in agreement with the assumption of perfect stationarity in the more detailed paper of Settnes and Bruus (see equations A1a to A1f of ref. \cite{Settnes2012}).

 Now, in the more general case of  a particle free to move (\textit{i.e.} when $\mathbf{F^*=0}$
 the particle movement is no longer exactly periodic in time so that
 a tiny incremental displacement $\delta(t)$ accumulate at every sound cycle.
 Consequently, the fluid momentum is no longer periodic and the  momentum  term  $\langle\frac{d}{dt}\int_{V(t)}\varrho\mathbf{v}dV\rangle $ is \textit{a priori}  not exactly zero
and the total radiation force differs from the Gor'kov expression. 
Let us estimate the order of magnitude of $\delta$ and subsequently of the associated correction onto the force.

 If we denote by $r_p(t)$ and $r^*_p(t)$  the particle displacement  respectively for a non suspended 
 and a suspended particle (in the presence  of $F^*$),
 we can write:
 \begin{equation}
 r_p(t)=r^*_p(t) + \delta(t).
 \end{equation}
 
 Let us first  estimate $\delta$.
 When pushed by the radiation force , the force acting on the particle is of the order $\varrho_f c^2 a^2 O({ \varepsilon^2+\alpha})$.
At large times (quasi stationary regime ) 
viscosity can no longer be neglected so that  the particle will reach a drift velocity of the order $F_{rad} / (\mu a) \sim c O(\varepsilon^{2+\alpha})$
(Stokes drag).
   Therefore, the difference  $\delta$ between a suspended  particle and a free one is of the order $\lambda O(\varepsilon^{2+\alpha})$.
The force on the particle writes (following the notations introduced in the manuscript)
\begin{equation}
\mathbf{F}(r_p(t)) =  \mathbf{F}(r_p^*(t) + \delta(t) )  = \mathbf{F}_a^{\text{loc}}  + \nabla (\tilde{\mathbf{F}_a} . (r_p^*+ \delta ))
\end{equation}

 In terms of forces, the difference between the
zero-mean displacement particle  and the free-to-move  particle 
yields a correction $\delta F$:
 \begin{equation}
\mathbf{\delta F} =  \nabla (\tilde{\mathbf{F_a}}) .  \delta 
 \end{equation}
which is of higher order.

In other words, the estimation of $\langle {\mathbf{F_a}} \rangle$ assuming periodicity does not significantly differ from its estimation for a free particle. Nevertheless, for a free particle,   $\langle {\mathbf{F_a}} \rangle= - \nabla U$ does not represent the total radiation force $\bar{\mathbf{F}}_{\text{rad}}$.

\section{SM2. Local term expansion}
\label{app:fa_loc}

In order to derive the $\varepsilon$ expansion of the local term $\mathbf{F}_a^{\text{loc}}$, we first insert the constant particle mass $m_p=\varrho_p(t)V_{p,in}(t)$ into Eq. 
(12) of the main paper: 

\begin{eqnarray}
\mathbf{F}_a^{\text{loc}}(t)&=&\frac{(\varrho_{p}-\varrho_{f})}{\varrho_{f0}} \boldsymbol{\nabla}p_{\textrm{in}} V_{p}\\
&=&m_p(1-\frac{\varrho_f}{\varrho_{p}})\frac{\boldsymbol{\nabla}p_{\textrm{in}}}{\varrho_{f0}}.
\label{eq:inst_rad_force_2}
\end{eqnarray}

We then successively get
\begin{eqnarray}
\mathbf{F}_a^{\text{loc}}(t)&=&m_p\left[\left(1-\frac{\varrho_f}{\varrho_p}\right)\frac{\boldsymbol{\nabla}p_{\textrm{in}}}{\varrho_{f0}}\right](\mathbf{r},t)\label{eq:fa_loc_der_1}\\
&=&m_p\left[\left(1-\frac{\varrho_{f0}}{\varrho_{p0}}-\frac{\varrho_{f0}}{\varrho_{p0}}(\kappa_{f0}-\kappa_{p0})p_{\textrm{in}}\right)\frac{\boldsymbol{\nabla}p_{\textrm{in}}}{\varrho_{f0}}\right](\mathbf{r},t)\nonumber\\
~\label{eq:fa_loc_der_2}\\
&=&-V_{p0}\left[-\frac{3}{2}\frac{f_2}{1-f_2}\boldsymbol{\nabla}p_{\textrm{in}}+\kappa_{f0}\frac{f_1}{2} \boldsymbol{\nabla}p^2_{\textrm{in}}\right](\mathbf{r},t)\label{eq:fa_loc_der_3},
\end{eqnarray}
passing from Eq. \ref{eq:fa_loc_der_2} to \ref{eq:fa_loc_der_3} using $(f_1,f_2)$ definition at Eqs. (7) and (8).

Moreover, we emphasize that between Eq.~\ref{eq:fa_loc_der_1} and  Eq. \ref{eq:fa_loc_der_2}
we used the postulate about the volume $V_{p,in}$ and $S_{p,in}$ as discussed in the main  article. More precisely, writing $\varrho_f\simeq\varrho_{f0}(1+\kappa_{f0}p_{\textrm{in}})$  and similarly $\varrho_p\simeq\varrho_{p0}(1+\kappa_{p0}p_{\textrm{in}}$) we consider explicitly the effect of the incident pressure, i.e., omitting the effect of $p_s$.

\section{SM3. Convective term expansion}
\label{app:fa_conv}

Starting from Eq. (21), we first have

\begin{eqnarray}
\mathbf{F}_{\text{conv}}(t)&=&V_{p_0}\frac{3}{2}f_{2}\int_{0}^{t}\boldsymbol{v}_{\textrm{in}}(\bold{r},t')dt'\cdot\boldsymbol{\nabla}\left(\boldsymbol{\nabla}p_{\textrm{in}}(\bold{r},t)\right)\label{eq:conv_term_dev2}\\
&=&V_{p0}\frac{3}{2}f_{2}\int_{0}^{t}\boldsymbol{v}_{\textrm{in}}(\bold{r},t')dt'\cdot\boldsymbol{\nabla}\left(-\varrho_{f0}\frac{\partial \boldsymbol{v}_{\textrm{in}}}{\partial t}(\bold{r},t)\right),\nonumber\\
~\label{eq:conv_term_dev3}
\end{eqnarray}
where we passed from Eq. \ref{eq:conv_term_dev2} to \ref{eq:conv_term_dev3} using the linearized Euler's equation.

Then, with respect to the mean contribution of the convective term, we successively get
\begin{eqnarray}
&\langle&\mathbf{F}_{\text{conv}}\rangle=\frac{1}{T}\int_{0}^{T}\mathbf{F}_{\text{conv}}(t)dt\label{eq:conv_term_dev5}\\
&=&-V_{p0}\varrho_{f0}\frac{3}{2}f_{2}\frac{1}{T}\int_{0}^{T}\int_{0}^{t}\boldsymbol{v}_{\textrm{in}}(\bold{r},t')dt'\cdot\frac{\partial}{\partial t}\boldsymbol{\nabla}\boldsymbol{v}_{\textrm{in}}(\bold{r},t)dt\nonumber\\
~\label{eq:conv_term_dev7}\\
&=&-V_{p0}\varrho_{f0}\frac{3}{2}f_{2}\frac{1}{T}\Bigg[\left[\int_{0}^{t}\boldsymbol{v}_{\textrm{in}}(\bold{r},t')dt'\cdot\boldsymbol{\nabla}\boldsymbol{v}_{\textrm{in}}(\bold{r},t)\right]_{0}^{T} \nonumber \\
&~&~~~~~~~~~~~~~~~~~~~~~-\int_{0}^{T}\left(\boldsymbol{v}_{\textrm{in}}\cdot\boldsymbol{\nabla}\boldsymbol{v}_{\textrm{in}}\right)(\bold{r},t')dt'\Bigg]\label{eq:conv_term_dev8}\\
&=&V_{p0}\varrho_{f0}\frac{3}{2}f_{2}\frac{1}{T}\int_{0}^{T}\left(\boldsymbol{v}_{\textrm{in}}\cdot\boldsymbol{\nabla}\boldsymbol{v}_{\textrm{in}}\right)(\bold{r},t')dt'\label{eq:conv_term_dev9}\\
&=&V_{p0}\varrho_{f0}\frac{3}{4}f_{2}\boldsymbol{\nabla}\left(\frac{1}{T}\int_{0}^{T}v^2_{\textrm{in}}(\bold{r},t')dt'\right)\label{eq:conv_term_dev11}\\
&=&V_{p0}\varrho_{f0}\frac{3}{4}f_{2}\boldsymbol{\nabla}\langle{v}^2_{\textrm{in}}\rangle(\bold{r})\label{eq:conv_term_dev12},
\end{eqnarray}
where Eq. \ref{eq:conv_term_dev5} transforms to \ref{eq:conv_term_dev7} from the gradient and time derivative linearities, Eq. \ref{eq:conv_term_dev7} to \ref{eq:conv_term_dev8} after an integration by parts, Eq. \ref{eq:conv_term_dev8} to \ref{eq:conv_term_dev9} as the incoming wave is such that $\frac{1}{T}\int_{0}^{T}\boldsymbol{v}_{\textrm{in}}(\bold{r},t')dt'=0$, and finally Eq. \ref{eq:conv_term_dev9} to \ref{eq:conv_term_dev11} through the gradient and integration operators linearities and taking into account $\boldsymbol{\nabla}\times\boldsymbol{v}_{\textrm{in}}=0$.\\
\indent 
We emphasize that  by  applying  a similar expansion as obtained for $\mathbf{F}_a(t)$ we can also calculate 
  \begin{eqnarray}
\mathbf{F}_b(t)=\langle\int_{S_{p,in}(t)}p_{\textrm{in}}\boldsymbol{dS}\rangle
\end{eqnarray}  which after a separation $\mathbf{F}_b(t)\simeq\mathbf{F}_b^{\text{loc}}(t)+\mathbf{F}_b^{\text{conv}}(t)$ similar to Eqs.~14-17 leads  to  
\begin{eqnarray}
\langle\mathbf{F}_b(t)\rangle\simeq V_{p0}\left(\frac{\kappa_{p0}}{2}\boldsymbol{\nabla} \langle p_{\textrm{in}}^2\rangle-\frac{1-f_2}{2}\varrho_{f0}\boldsymbol{\nabla} \langle v_{\textrm{in}}^2\rangle\right).\nonumber \\
\end{eqnarray}
This formula is important since it shows that even in the case of $f_1=f_2=0$, for a standing wave, the contribution  $\langle\int_{S_{p,in}(t)}p_{\textrm{in}}\boldsymbol{dS}\rangle$ doesn't  vanish.  More precisely in this regime we have 
\begin{eqnarray}
\langle\mathbf{F}_b(t)\rangle\simeq V_{f0}\left(\frac{\kappa_{f0}}{2}\boldsymbol{\nabla} \langle p_{\textrm{in}}^2\rangle-\frac{1}{2}\varrho_{f0}\boldsymbol{\nabla} \langle v_{\textrm{in}}^2\rangle\right).\nonumber \\ \simeq V_{f0}\boldsymbol{\nabla} \langle p_{\textrm{in}}^{(2)}\rangle
\end{eqnarray} where appears the second order acoustic pressure $p_{\textrm{in}}^{(2)}=\frac{\kappa_{f0}}{2} p_{\textrm{in}}^2-\frac{1}{2}\varrho_{f0} v_{\textrm{in}}^2$ in the fluid.\\
 For a standing wave with $p_{in}=A\cos{(\omega t)}\cos{(\omega x/c_{f0})}$, $\boldsymbol{v}_{\textrm{in}}=\mathbf{\hat{x}}\frac{A}{c\varrho_{f0}}\sin{(\omega t)}\sin{(\omega x/c_{f0})}$ we have in particular
 $ \langle p_{\textrm{in}}^{(2)}\rangle=\frac{\kappa_{f0}A^2}{4}\cos{(2\omega x/c_{f0})}$ and thus 
\begin{eqnarray}
\langle\mathbf{F}_b(t)\rangle\simeq -\mathbf{\hat{x}}V_{f0}\frac{2\omega}{c_{f0}}\frac{\kappa_{f0}A^2}{4}\sin{(2\omega x/c_{f0})}.
\end{eqnarray}
 In the main text this motivated the introduction of the coefficient $\beta(t)$ in Eq.~10 and all the subsequent developments. 

\bibliography{biblio_radiation_force_2018}